\documentclass[10pt,letterpaper,twocolumn]{article} 

\usepackage{ol2}
\usepackage[draft]{hyperref}
\usepackage{amsmath}

\begin{document}

\twocolumn[ 

\title{Coherent virtual absorption for discretized light}


\author{Stefano Longhi}

\address{Dipartimento di Fisica, Politecnico di Milano and Istituto di Fotonica e Nanotecnologie del Consiglio Nazionale delle Ricerche, Piazza L. da Vinci 32, I-20133 Milano, Italy (stefano.longhi@polimi.it)}

\begin{abstract}
Coherent virtual absorption (CVA) is a recently-introduced phenomenon for which exponentially growing waves incident onto a conservative optical medium are neither reflected nor transmitted, at least transiently. CVA has been  associated to complex zeros of the scattering matrix and can be regarded as the time reversal of the decay process of a quasi-mode sustained by the optical medium.  Here we consider CVA for discretized light transport in coupled resonator optical waveguides or waveguide arrays and show that a distinct kind of CVA, which is not related to complex zero excitation of quasi-modes, can be observed. This result suggests that scattering matrix analysis can not fully capture CVA phenomena. 
\end{abstract}

 ] 

The ability of tailoring light absorption in a medium exploiting interference effects has sparked much interest in the past recent years with a wealth of potential applications ranging from  optical sensing to optical modulation, optical switching, polarization control, etc. \cite{r3,r4,r5,r6,r7,r8,r10,r13,r14,r16,r19,r20,r21,r22,r23,r25,r27}; for a recent review see \cite{r28}. Coherent perfect absorption (CPA) refers to the possibility for a dissipative medium to fully absorb suitable patterns of incoming waves; it can be regarded as the time reversal of  lasing \cite{r4}. In a multiport linear system, incoming and outgoing fields are related by the scattering matrix $S$. CPA is realized when an eigenvalue of $S$ vanishes at some real frequency. In a recent work \cite{r27}, Baranov and collaborators introduced the concept of coherent virtual absorption (CVA) in lossless reciprocal media, which is based on the excitation of the medium with suitable incoming waves with complex frequency. While CPA modes are associated to zeros of the scattering matrix on the real axis, CVA corresponds to {\it complex } zeros of the scattering matrix, i.e. existence of quasi-bound optical modes. Unlike conventional CPA, in CVA the excitation fields have an amplitude that increases exponentially in time and space, similar to Gamow$^{\prime}$s states introduced in earlier theories of quantum mechanical decay \cite{r29}. CVA can be viewed as the time reversal process \cite{r29bis,r29tris} of the light decay process in a high-quality passive resonator \cite{r29tris}.\\
\begin{figure}[htb]
\centerline{\includegraphics[width=8.4cm]{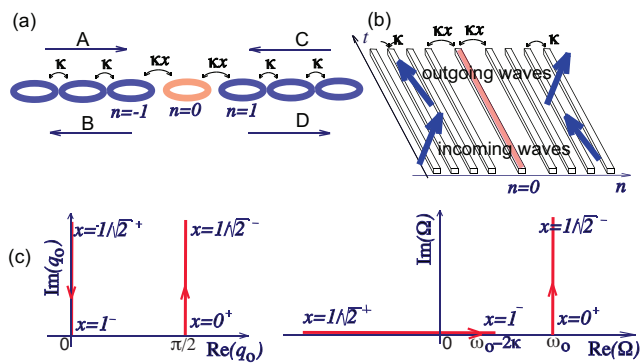}} \caption{ \small
(Color online) (a,b) Scattering for discretized light in (a) a CROW with a defective resonator at site $n=0$, and (b) an array of dielectric optical waveguides with a defect at site $n=0$. (c) Behavior of the zero $q=q_0$ of the scattering matrix determinant versus $x$ in complex plane for a lossless CROW with $\gamma=\delta \omega_0=0$ (left panel) and corresponding behavior of the frequency $\Omega(q_0)$ (right panel).}
\end{figure}
In this Letter we reveal a distinct form of coherent virtual absorption, which is neither related to complex zero eigenvalues nor does it strictly require existence of quasi-bound optical eigenmodes. Specifically, we consider
transport  of discretized light in coupled resonator optical waveguides (CROW) \cite{r30,r31} or in evanescently-coupled optical waveguide arrays \cite{r34,r35,r36}, and show that a kind of virtual absorption can be observed even in structures that do not sustain any resonance mode, i.e. beyond the conditions considered in previous work \cite{r27}. Such a kind of CVA stems from the property that specially-designed  waveforms, synthesized using time reversal symmetry, can show half-space  discrete self-focusing on the lattice.\\
We consider light transport in a linear chain of CROW \cite{r30,r31} [Fig.1(a)] or, similarly, spatial propagation of light in arrays of evanescently coupled optical waveguides \cite{r34,r35,r36} [Fig.1(b)]. For the sake of definiteness, we will refer mostly to the former physical system.  The CROW is composed by a chain resonators with the same resonance frequency $\omega_0$, which are evanescently-coupled with a coupling constant $\kappa$. A defective resonator, with detuned resonance frequency $\omega_0+\delta \omega_0$, is placed at site $n=0$, and coupled to its neighboring resonators with a modified coupling constant $x \kappa$, with $x \leq 1$; see Fig.1(a). The defective resonator could be lossy, with a loss rate $\gamma$ ($\gamma=0$ in the conservative limit). Light transport in the CROW is described by coupled-mode equations \cite{r30,r31}
\begin{eqnarray}
i \frac{dc_n}{dt} & = & \omega_0 c_n -\kappa (c_{n+1}+c_{n-1}) \;\;\; |n| \geq 2 \nonumber \\
i \frac{dc_{\pm1}}{dt} & = &\omega_0 c_{\pm 1} -\kappa c_{\pm2}-x \kappa c_0 \\
i \frac{dc_0}{dt} & = & (\omega_0+\delta \omega_0-i \gamma) c_0-x \kappa (c_1+c_{-1}) \nonumber
\end{eqnarray}
where $c_n(t)$ is the field amplitude of the resonator mode at site $n$. The scattering properties of the defective resonator are embodied in the $2 \times $2 scattering matrix 
\begin{equation}
S= \left(
\begin{array}{cc}
r & t \\
t & r
\end{array}
\right)
\end{equation}
where $r=r(q)$ and $t=t(q)$ are the spectral reflection and transmission amplitudes, respectively. These are determined from the solutions to Eq.(1) of the form
\begin{eqnarray}
c_n(t)= \left\{
\begin{array}{cc}
\left[ A \exp(iqn) +B \exp(-iqn) \right] \exp(-i \Omega t) & n \leq -1 \\
E \exp(-i \Omega t) & n=0 \\
\left[ C \exp(-iqn)+D \exp(iqn) \right] \exp(-i \Omega t) & n \geq 1
\end{array}
\right.
\end{eqnarray}
where $(A,C)$ and $(B,D)$ are the amplitudes of incoming and outgoing waves, respectively [Fig.1(a)], $q$ is the Bloch wave number with $0 \leq {\rm Re}(q) \leq \pi$, and 
\begin{equation}
\Omega(q)=\omega_0-2 \kappa \cos q
\end{equation}
is the lattice dispersion relation of the homogeneous lattice \cite{r36}. The scattering matrix relates the amplitudes of outgoing and incoming waves according to $(B,D)^T=S (A,C)^T$.
The explicit form of $r(q)$ and $t(q)$ read
\begin{eqnarray}
r(q)=\frac{\rho / \kappa - 2(x^2-1) \cos q}{2 x^2\exp(iq)- 2 \cos q - \rho / \kappa} \\
 t(q)=\frac{2i x^2 \sin q }{2 x^2\exp(iq)- 2 \cos q - \rho / \kappa} 
\end{eqnarray}
where we have set $\rho \equiv \delta \omega_0-i \gamma$. Note that, for a real Bloch wave number $q$, the frequency $\Omega(q)$ is real and the scattered states (3) are stationary and propagative waves.\\
Like for light scattering in continuous media, CPA and CVA based on real and complex zeros of the scattering matrix  can be found for discretized light on a lattice. 
The condition ${\rm det} (S)=0$, corresponding to a zero eigenvalue of $S$, implies $r=-t$ or $r=t$. Only the former condition $r=-t$, corresponding to symmetric mode excitation $A=C$, is relevant and yields the following  equation for the Bloch wave number $q$
\begin{equation}
(x^2-1) \cos q-i x^2 \sin q= (\delta \omega_0-i \gamma)/ ( 2 \kappa)
\end{equation}
For CPA, we require $q$ to be real. Like in other CPA systems  \cite{r4}, this implies some dissipation $\gamma$ in the medium and a constraint between frequency detuning $\delta \omega_0$ and dissipation $\gamma$.
\begin{figure}[htb]
\centerline{\includegraphics[width=8.4cm]{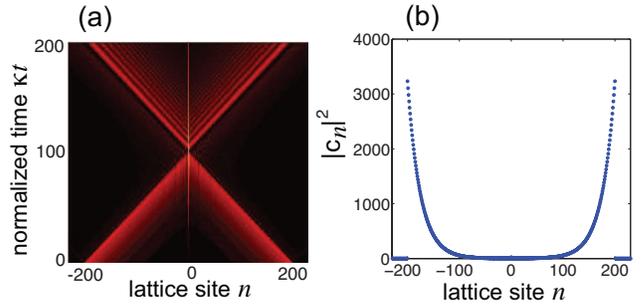}} \caption{ \small
(Color online) Coherent virtual absorption in a lossless CROW excited by incoming Gamow$^{\prime}$s waves, truncated at $|n|>200$. Parameter values are $\delta \omega_0=\gamma=0$, $x=0.2$. (a) Snapshot of $|c_n(t)|$ on a pseudo color map, clearly showing perfect energy loading of the defective resonator (site $n=0$) at the focusing time $t=T=$. For $t>T$, the energy is released in a time reversal process. (b) Behavior of initial excitation $|c_n(0)|^2$  (truncated Gamow$^{\prime}$s wave).}
\end{figure}
In a dissipationless structure, i.e. for $\gamma=0$, the condition ${\rm det}(S)=0$ can be satisfied for a complex Bloch wave number $q=q_0$, corresponding to incoming (Gamow$^{\prime}$s-like) waves with an exponentially increasing amplitude in time. For the sake of simplicity, let us assume $\delta \omega_0=0$ and $x \leq 1$, so that the defect does not sustain any bound state. A resonance (quasi-bound) state, with lifetime determined by the imaginary part of $\Omega(q_0)$, is found as the complex root of Eq.(7) with $\gamma=\delta \omega_0=0$
\begin{figure*}[htb]
\centerline{\includegraphics[width=18.3cm]{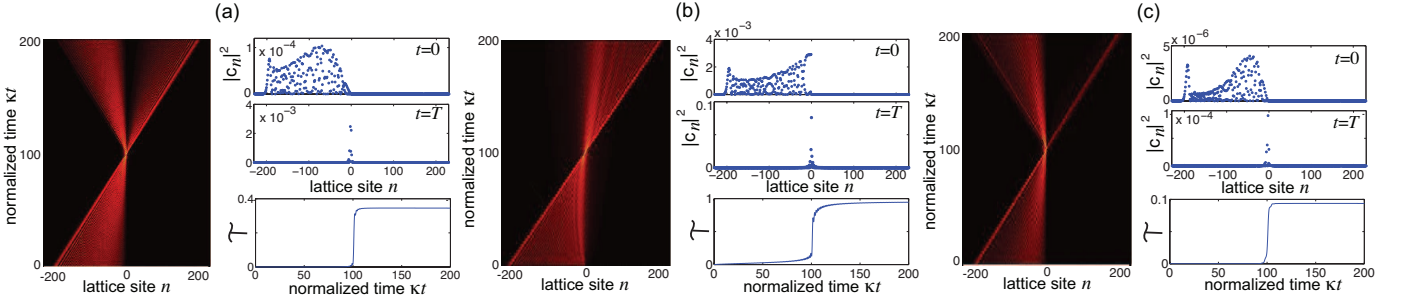}} \caption{ \small
(Color online) Virtual absorption for one-side excitation in a lossless CROW for $\delta \omega_0=0$ and for a few values of $x$: (a) $x=1/ \sqrt{2}$ (phase transition point), (b) $x=1$ (above the phase transition point, uniform CROW), and (c) $x=0.4$  (below the phase transition point). Left panels show the temporal evolution of mode amplitudes $|c_n(t)|$. Initial condition is defined by Eqs.(12) and (13) with $T=100/ \kappa$ and $\beta=0.015$ in (a), $\beta=0.005$ in (b) and $\beta=0.025$ in (c). The detailed behavior of $|c_n(t)|^2$ at initial time $t=0$ and at the focusing time $t=T$ is shown on the right panel (upper two plots), whereas the time behavior of the transmittance $\mathcal{T}$ is shown in the lower plot. $\mathcal{T}$ is defined by Eq.(14) with $n_0=2$.}
\end{figure*}
\begin{equation}
q_0=\pi /2 - (i/2) {\rm ln}(1-2x^2).
\end{equation}
 Note that for $x \ll 1$ one has $q_0 \simeq  \pi/2+i x^2$, and the complex energy of the resonance mode is given by $\Omega \simeq\omega_0 + 2 i \kappa  x^2$. In this limit, two incoming waves, exponential growing in time $t$ and space $n$ according to
 \begin{equation}
 c_n(t) \simeq \exp [-i(\pi/2)|n|+x^2 |n|+2 \kappa x^2 t-i \omega_0 t]
 \end{equation}
($n \neq 0$),  realize CVA, in which the energy of incoming waves is fully stored in the defective resonator. Figure 2 shows an example of CVA based on complex zero excitation, in which the exponential waves at initial time are  truncated at $|n|=200$ [Fig.2(b)]. This is precisely the scenario considered in Ref.\cite{r27}, and therefore it will not be further analyzed.\\
Here we wish to suggest a different route toward virtual absorption, which is not associated to the existence of resonance modes.  To this aim, let us consider the behavior of the Bloch wave number $q_0$ [Eq.(8)] in complex plane,  which is shown in Fig.1(c) together with the corresponding behavior of the frequency $\Omega=\omega_0-2 \kappa \cos q_0$. Clearly, a phase transition is found at $x=1/ \sqrt{2}$. 
For $x< 1 / \sqrt{2}$, ${\rm Re}(q_0)=\pi/2$ is independent of $x$ while ${\rm Im}(q_0)$ increases as $x$ increases, corresponding to lowering of the $Q$-factor (lifetime) of the quasi-mode. At $x=1/ \sqrt{2}$, $q_0$ goes to infinity, i.e. ${\rm det}(S)$ does not vanish nowhere in complex $q$ plane and the quasi-mode disappears.  For $1/ \sqrt{2} < x \leq 1$, $q_0$ is purely imaginary and the frequency $\Omega$ becomes real, indicating that the solution (3) with $B=D=0$ is stationary in time but exponentially growing in space (anti-evanescent wave), and thus unphysical.\\
 A natural question then arises: while for $x< 1 / \sqrt{2}$ the structure sustains a quasi-bound mode and CVA can be observed by complex zero eigenvalue excitation, what happens when crossing the phase transition point at $x=1/ \sqrt{2}$? Can virtual absorption be found for $x> 1 / \sqrt{2}$? The main result of this Letter is that virtual absorption does not strictly require the existence of quasi bound modes. Exploiting the very general concept of time-reversal symmetry, one can synthesize half-space  self-focusing wave packet distributions  such that, for a finite (yet arbitrarily long) time interval, an initially broad wave, localized at $n<0$ sites, shrinks  and stores energy in few lattice sites around $n=0$, with negligible transmission in the $|n|>1$ half space.\\ 
 To gain some insights into the synthesis method, let us indicate by $U_{n,m}(t)$ the propagator of Eq.(1) with $\delta \omega_0=\gamma=0$, so that 
 \begin{equation}
 c_n(t)=\sum_m U_{n,m}(t) c_m(0) \exp(-i \omega_0 t).
 \end{equation}
  The propagator $U_{n,m}(t)$ can be calculated in a closed form for special values of $x$, namely for $x=1$ (homogeneous lattice) and for $x= 1 / \sqrt{2}$, i.e. at the phase transition point. For $x=1$, one has $U_{n,m}(t)=i^{n-m} J_{n-m}(2 \kappa t)$ \cite{r34,r35,r36}, while for $x= 1 /\sqrt{2}$ the expression of $U_{n,m}(t)$ is more involved and not given here for the sake of brevity (see \cite{r38}). For other values of $x$, $U_{n,m}(t)$ can be computed numerically.  CVA based on time reversal can be exemplified by considering two simple cases.\\
  (i) As a first example, let us assume $c_n(0)=\delta_{n,0}$. After some time $t=T$, a broad symmetric waveform is found as a result of wave spreading in the lattice, which is given by $c_{n}(T)=U_{n,0}(T)$ with $|c_{-n}(T)|^2=|c_n(T)|^2$.  Its time reversal corresponds to a symmetric self-focusing wave. In fact, assuming as an initial condition $c_n(0)=U_{n,0}^{*}(T)$, complex conjugation corresponds to time reversal and the spatially-broad wave distribution shrinks and completely stores energy at the $n=0$ lattice site at the focusing time $t=T$, i.e. $|c_n(T)|=\delta_{n,0}$. Note that this phenomenon of self-focusing as time reversal of discrete spreading in the lattice is a very general one and is not related to complex zero eigenmodes.\\ 
  (ii) As a second example, let us consider excitation of the site $n=0$ {\it and} of a few other sites around $n=0$ with tailored phases and amplitudes. In this case, wave spreading in the lattice becomes asymmetric, and its time reversal thus corresponds to asymmetric CVA, i.e. CVA with unbalanced wave amplitudes from the two sides of the $n=0$ site. For example, let us consider the simplest case $x=1$ and let us assume $c_n(0)=\delta_{n,0}+i\Delta ( \delta_{n,1}+\delta_{n,-1})$; one then readily obtains 
\begin{equation}
c_n(T)=i^n J_n( 2 \kappa t) \left(1 + \frac{n \Delta}{\kappa T} \right).
\end{equation}
Note that the waveform broadens as $\kappa T$ increases and the symmetry $|c_{-n}|^2=|c_n|^2$ is broken for a non-vanishing value of $\Delta$. Since the Bessel $J_n$ function has its maxima at $n \simeq \pm 2 \kappa T$ for $\kappa T \gg 1$, by choosing $\Delta=-1/2$ the distribution $|c_n(T)|^2$ turns out to be strongly asymmetric around $n=0$, with the major excitation in the $n<0$ half lattice. Time reversal thus corresponds to a self-focusing wave with dominant excitation in the left half-space of the lattice. If more than three waveguides with properly-tailored amplitudes and phases are excited at $t=0$, the pattern $|c_n(T)|^2$ can be entirely localized in the half space $n \leq 0$, so that its time reversal would correspond to  CVA with unidirectional excitation of the lattice. This is a distinctive feature as compared to CVA based on complex zeros of the scattering matrix eigenvalues, where bidirectional excitation is required [see Eq.(9)].\par 
\begin{figure}[htb]
\centerline{\includegraphics[width=8.4cm]{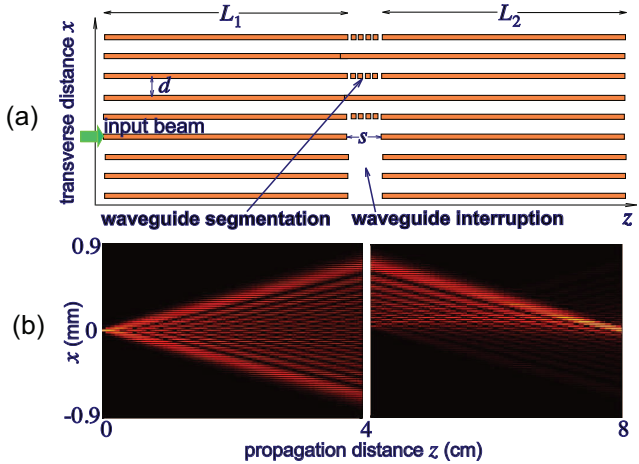}} \caption{ \small
(Color online) (a) Schematic of a waveguide array for an experimental observation of CVA. Single waveguide excitation at input $z=0$ produces symmetry diffraction pattern in the first section of the array. At $a=L_1$ waveguides in the $x<0$ half space are interrupted, preventing further light propagation, whereas waveguides in the $x>0$ half-space with odd index $n$ are segmented realize alternating $\pi$ phase shifts. (b) Simulated optical beam propagation in the waveguide array based on coupled mode equations.}
\end{figure}
To realize CVA with one-side excitation, let us assume the initial condition 
\begin{equation}
c_{n}(0)=U_{n,0}^*(T)H_n
\end{equation}
 with envelope $H_{n}$ defined by
\begin{equation}
H_n= \left\{
\begin{array}{cc}
\exp( \beta n) & n \leq -1 \\
0 & n \geq 0
\end{array}
\right.
\end{equation}
and $0<\beta  \ll 1$. The waveform (12) is localized in the $n<0$ half-space and is obtained from the deformation and truncation of the time reversal of the impulse response $U_{n,0}$ of the lattice. Extended numerical simulations indicate that the waveform (12) undergoes half-space transient self focusing in the time interval $(0,T)$, with $c_n(T)$ being non vanishing for few sites around $n=0$ (focusing condition). The instantaneous transmittance $\mathcal{T}(t)$, defined by 
\begin{equation}
\mathcal{T}(t) \equiv \frac{\sum_{n=n_0}^{\infty} |c_n(t)|^2}{ \sum_{n=-\infty}^{\infty} |c_n(0)|^2}
\end{equation}
  ($n_0>0$ is some small integer), turns out to be negligible for $t<T$, showing an abrupt transition to a non-negligible value at $t>T$. This result is a rather general one and turns out to be rather independent of the small parameter $\beta$, which can be used to flatten the wave excitation amplitudes over the lattice sites at initial time. Examples of self-focusing  waveforms, corresponding to virtual absorption for unidirectional wave excitation, are shown in Fig.3 for a few values of $x$. For times smaller than the focusing time $t=T$, the initially broadened wave packet shrinks toward the $n=0$ site but is not able to cross this site, as if virtual absorption would occur (actually energy is increasingly stored in a few sites around $n=0$). However, at times larger than $T$ wave scattering (either transmission or reflection) is clearly visible. Interestingly, such a method to synthesize discrete wave packets  can be applied for $x<1 / \sqrt{2}$ as well, i.e. below the phase transition point [Fig.3(c)]. In this case virtual absorption is clearly not related to the complex zero of the scattering matrix eigenvalue: in fact, the wave packet excitation greatly deviates from the exponential Gamow$^{\prime}$s state, and excitation is unidirectional rather than bidirectional [compare Fig.2 and Fig.3(c)].\par
  {\it Physical observation of virtual absorption without complex zero excitation.}  An experimentally accessible system for the observation of CVA of discretized light is provide by spatial light propagation in waveguide arrays \cite{r34,r35,r36}, where the propagation distance $z$ along the array plays the same tole as time $t$ in a CROW. A suggested experimental setup is shown in Fig.4(a). A waveguide array with homogeneous waveguide spacing $d$ [corresponding to $x=1$ and $\delta \omega_0=0$ as in Fig.3(b)] comprises two sections, of length $L_1$ and $L_2 \geq L_1$. The first section is used to synthesize the waveform (12), which then propagates in the second section to realize CVA. The array is excited at the $n=0$ waveguide and discrete diffraction occurs from $z=0$ up to $z=L_1$. At $z=L_1$ half waveguides of the array, at sites $n \leq 0$, are interrupted for a short length $s$, preventing further propagation of light in the second section of the array.  $s$ is chosen of the order (or smaller) than the coupling length $\sim 1 / \kappa$, yet sufficiently long to prevent light re-coupling after the interruption. In the short distance $s$, the other half waveguides of the array are alternately segmented to realize $\pi$ phase slips between adjacent guides \cite{r40,r41}. The phase slips realize the analog of time reversal \cite{r41,r41bis}. The optical pattern that propagates in the second section of the array is thus the waveform (12) with $\beta=0$. Figure 4(b) shows an example of simulated discrete light propagation in a 8-cm-long array comprising 80 waveguides, fabricated by fs-laser-writing in fused silica \cite{r35}, for a lattice period  $d=24 \; \mu$m, corresponding to a coupling constant $\kappa \simeq 0.4 \; {\rm mm}^{-1}$  \cite{r42}, and segmentation/interruption length $ \sim 2.6$ mm \cite{r40}. Clearly, in the second array section the discrete beam undergoes half self-focusing, realizing CVA without complex zero eigenmodes. Fluorescence imaging techniques \cite{r35,r40} can be used to experimentally reproduce propagation maps of Fig.4(b).\\ 
  In this work the phenomenon of coherent virtual absorption, recently introduced in \cite{r27}, has been investigated for discretized light transport. The main conclusion is that virtual absorption does not necessarily require complex zero excitation of quasi-modes. This result suggests that scattering matrix analysis does not fully capture the entire physics of the phenomenon and motivates further theoretical and experimental research,  opening up some major questions. For example, can virtual absorption without complex zero eigenvalue excitation be observed for light waves in continuous media? Can transient virtual absorption be controlled by exploiting material nonlinearity?

\end{document}